# Band structure and inter-tube optical transitions in double-walled carbon nanotubes

D.V. Chalin and S.B. Rochal

*Faculty of Physics, Southern Federal University, 5 Zorge str., 344090, Rostov-on-Don, Russia*

Usually, in optical spectra of double-walled carbon nanotubes (DWCNTs) weak van der Waals coupling between the layers leads only to a small shift of transition energies with respect to their values in pristine single-walled nanotubes. However, recent results have shown that the Rayleigh spectrum of the DWCNT (12,11)@(17,16) contains additional peaks. Using the tight binding approximation, we demonstrate that in specific DWCNTs the interlayer coupling can slightly modify the band structure of pristine nanotubes in such a way that the unconventional inter-tube electronic transitions become possible and additional peaks in the DWCNT optical spectrum appear. Using the known experimental data on 118 optical transitions in DWCNTs, in addition to the recently published case we reveal six more DWCNTs with inter-tube transitions and obtain geometrical selection rules permitting them. In few dozens of DWCNTs our approach yields the energies of electronic transitions close to the experimentally observed ones and may be useful for structural identification of this-type nanotubes.

## I. INTRODUCTION

In recent years graphene-based and analogous bilayer systems have attracted increasing attention due to their unique properties. In these modern materials, close connection between the structural organization that can demonstrate very beautiful moiré patterns[1,2] and electronic band structure opens up new possibilities for altering their electronic properties. For example, by varying the angle between the layers in twisted bilayer graphene (TBLG), one can change the position of the Van Hove singularities[3,4] (VHS) in the density of electronic states of the TBLG.[1,3,4] When the VHS is close to the Fermi energy, electronic instabilities arise in the bilayer superlattice, which can result in the appearance of new states such as superconducting[5,6] and Mott insulating ones.[6,7] TBLG is also one of the few systems where it was possible to observe the fractal structure of electronic spectrum in the magnetic field, called the Hofstadter's butterfly.[8] Despite the great progress made in the study of planar incommensurate systems, such as TBLG, graphene on boron nitride[9,10] and $MoS_2$ structures,[11,12] the electronic properties of DWCNTs remain poorly studied, and the number of papers devoted to the theoretical side of this problem hardly exceeds a dozen. All structurally identified double-walled carbon nanotubes (DWCNTs) investigated so far are incommensurate.[13–17] The incommensurability together with the curved geometry and structural diversity of DWCNTs can result in rather interesting electronic properties useful for various applications.[18,19]

Usually, the DWCNT optical spectrum is described as a simple superposition of the spectra of single-walled carbon nanotubes (SWCNTs) forming the DWCNT, with a small shift of spectral peaks caused by weak van der Waals interaction between the inner and outer layers.[14–16,20] Using this idea, in order to describe the electronic transitions in incommensurate DWCNTs, the authors of the pioneer paper[20] proposed a theoretical model based on perturbation theory. Let us note that virtually all previous theoretical studies considered only commensurate DWCNTs,[21–27] and the model proposed in Ref. 20 was the first one explaining the optical spectra of real DWCNTs. An alternative approach[28] allows one to calculate a small region of the band structure of incommensurate DWCNT. According to Ref. 28, if the nanotube has some certain structure, a global alteration of electronic bands occurs, and the resulting optical spectrum is far from a simple sum of the SWCNT spectra. The authors[28] argue that for such spectrum rearrangement, one of two conditions must be satisfied: either the chiral vectors directions for the inner and outer SWCNTs are close to each other, or the vectors difference is parallel to the armchair direction. For the first time an optical spectrum rearrangement has been experimentally observed in the DWCNT (12,11)@(17,16) only few month ago.[29] The measured Rayleigh spectrum besides the optical transitions originating from the inner and outer SWCNTs contained a few additional ones. The authors[29] explain the rearrangement of the band structure within the framework of the approach,[28] since both inner and outer tubes are nearly-armchair and their chirality angles are approximately equal.

In this Article, developing and revising the previous works, we introduce a concept of inter-tube electronic transitions and explain the mechanism of this phenomenon. As we demonstrate, the weak van der Waals interlayer coupling in specific DWCNTs can change the band structure of pristine SWCNTs in such a way that the electronic transition between the bands corresponding to different nanotubes becomes possible. Denying the global reconstruction of the band structure, we point out the origin of each band participating in the transition. The proposed theory uses the nearest neighbor tight-binding approximation[30,31] (NN TBA), which makes it a simple and convenient tool for analyzing the band structure of DWCNTs. Geometric selection rules permitting the inter-tube transitions are found. The rules obtained are not as strict as those formulated earlier[28] and allow the unconventional electronic transitions in DWCNTs with much greater structural diversity. Particularly, we detect the inter-tube transitions in the previously published optical spectra of DWCNTs (12,12)@(21,13)[20] and (10,6)@(14,13)[15], which violate the conditions.[28,29] In addition, the developed theory can also be successfully applied to describe conventional in-tube electronic transitions in DWCNTs. Comparing to Ref. 20, the obtained expressions for energy dispersions contain additional terms, which correspond to cross-band interaction and increase the accuracy of calculations. Totally, using the known data,[15,16,20] we have analyzed and successfully explained 118 conventional in-tube electronic transitions in 32 DWCNTs.

The rest of this Article is organized as follows. In the next section we reanalyze the theory[20] and present essentially more compact and more accurate expressions to calculate the energies of optical transitions. We also introduce the effective Hamiltonian which we use to analyze the band structure of incommensurate DWCNTs. In sections III and IV we present the concept of inter-tube electronic transition and justify it using the known experimental data. Section III mostly regards the peculiarities of the band structure in considered DWCNTs, while in Section IV we discuss and calculate dipole matrix elements for inter-tube transitions. In Discussion we compare the developed theory with the previous ones. The paper ends with Conclusion.

## II. RELATION BETWEEN THE DWCNT BAND STRUCTURE AND THE ONES OF PRISTINE NANOTUBES

In the framework of the nearest-neighbor tight binding model[30–33] the Hamiltonian of an individual SWCNT can be written as

$$H = \begin{pmatrix} 0 & f(\boldsymbol{q}) \\ f^*(\boldsymbol{q}) & 0 \end{pmatrix}, \quad (1)$$

where $f(\boldsymbol{q}) = \gamma\{\exp[-i(\boldsymbol{a}_1 + \boldsymbol{a}_2)/3 \cdot \boldsymbol{q}] + \exp[i(2\boldsymbol{a}_1 - \boldsymbol{a}_2)/3 \cdot \boldsymbol{q}] + \exp[i(-\boldsymbol{a}_1 + 2\boldsymbol{a}_2)/3 \cdot \boldsymbol{q}]\}$ is the matrix element describing the interaction between the graphene sublattices $A$ and $B$, $\gamma$ is the hopping coefficient; $\boldsymbol{a}_1$ and $\boldsymbol{a}_2$ are the graphene's primitive translations expressed in a cylindrical coordinate system of SWCNT, where the first and second components correspond to the projections on the circumference and the longitudinal axis of the tube, respectively (see Appendix A); $\boldsymbol{q} = (\mu, k)$ is the wave vector. Its first integer component $\mu$ numbers the cutting lines and the second component $k$ is the wave vector projection along them.[20,34] Eigen energies of the Hamiltonian (1) read $E^{\pm} = \pm|f(\mu,k)|$, while eigen vectors can be found as

$$|\psi^{\pm}\rangle = |\psi_A\rangle \pm e^{i\varphi}|\psi_B\rangle, \quad (2)$$

where $|\psi_A\rangle$ and $|\psi_B\rangle$ are the Bloch wave functions (WFs) of the sublattices[20,31,33], the positive and negative signs in Eq. (2) correspond to the states in the conduction band (CB) and the valence band (VB), respectively, $\varphi = \text{Arg}(|f|/f)$ is the phase shift between the sublattices. Thus, the electronic states in SWCNTs can be classified using the vector $(\mu, k)$ and the number $\sigma = \pm 1$, indexing the conduction (+1) and valence (-1) bands.

This formalism can also be generalized for the case of DWCNT. As shown in the paper,[20] in a DWCNT an electronic state of one tube can strongly interact with several states of the other tube. However, simplifying the problem, we first consider the interaction between two electronic states $(\boldsymbol{q},\sigma)$ and $(\boldsymbol{q}',\sigma')$ of the inner and outer SWCNTs, respectively. Let us write, using the matrix representation, the effective DWCNT Hamiltonian suitable to describe such "pair" interaction as:

$$\boldsymbol{H}_{DW} = \begin{pmatrix} \boldsymbol{H}_{in} & \boldsymbol{V} \\ \boldsymbol{V}^* & \boldsymbol{H}_{out} \end{pmatrix}, \quad (3)$$

where $\boldsymbol{H}_{in}$ and $\boldsymbol{H}_{out}$ are the Hamiltonians of the inner and outer SWCNTs with the form (1), $\boldsymbol{V}$ is the interlayer coupling matrix:

$$\boldsymbol{V} = \begin{pmatrix} V_{A,A'} & V_{A,B'} \\ V_{B,A'} & V_{B,B'} \end{pmatrix}. \quad (4)$$

In Eq. (4) $V_{\alpha,\beta} = \langle \psi_{(\alpha)}^{in}|\hat{V}|\psi_{(\beta)}^{out}\rangle$ are the matrix elements of the interlayer coupling operator $\hat{V}$, describing the interaction between the sublattices of the inner ($\alpha = A, B$) and outer ($\beta = A', B'$) nanotubes.

Following Ref. 20, these quantities can be written as

$$V_{\alpha,\beta} = \frac{1}{\sqrt{N_1 N_2}} \sum_{j,l} e^{-i[\boldsymbol{q}\cdot\boldsymbol{R}_j(\alpha) - \boldsymbol{q}'\cdot\boldsymbol{R}'_l(\beta)]} u(j,l), \quad (5)$$

where $\boldsymbol{R}_j(\alpha) = (\theta_j, z_j)$ and $\boldsymbol{R}'_l(\beta) = (\theta'_l, z'_l)$ are the cylindrical coordinates of atoms in the sublattices $\alpha$ and $\beta$, respectively; $N_1$ and $N_2$ are equal to the number of atoms in these sublattices; $u(j,l)$ is the interatomic matrix element between $p$-orbitals localized near atoms with the numbers $j$ and $l$. This integral depends only on the linear distance between the sites of the atoms.[20,23]

In incommensurate DWCNTs when we consider two arbitrarily electronic states $(\boldsymbol{q},\sigma)$ and $(\boldsymbol{q}',\sigma')$ of the inner and outer tubes, the matrix element $V_{\alpha,\beta}$, being averaged over spatial coordinates, vanishes due to oscillating phase. However, for strongly coupled electronic states this is not the case.[20,28] In particular, the strong coupling takes place for the modes with $\boldsymbol{q}' = \boldsymbol{q}$. Let us note that the coupling strength decays exponentially with increasing distance between $\boldsymbol{q}$ and $\Gamma$ points.[20]

As we show in Appendix B due to incommensurability of inner and outer nanotubes matrix elements $V_{\alpha,\beta}$ are practically real and independent on indices $\alpha$ and $\beta$. Accordingly, omitting the indices of sublattices and taking into account the properties considered above, we can rewrite the element $V_{\alpha,\beta}$ as:

$$h = \frac{1}{\sqrt{N_1 N_2}} \sum_{j,l} \cos[\boldsymbol{q}\cdot(\boldsymbol{R}_j - \boldsymbol{R}'_l)] u(j,l) \quad (6)$$

To calculate the $h$ value, the function $u(j,l)$ is chosen as $u = \gamma_c \exp(-r/\lambda)$, where $\gamma_c$ is the interlayer interaction strength, $r$ is the distance between the atom sites $j$ and $l$;[21,23] $\lambda = 0.045\ nm$ is the characteristic wavelength.[23] Our numerical analysis shows that the length of a DWCNT in calculations can be limited to 100 nm.

Thus, the effective Hamiltonian (3) can be simplified as

$$\boldsymbol{H}_{DW} = \begin{pmatrix} 0 & f_{in} & h & h \\ f_{in}^* & 0 & h & h \\ h & h & 0 & f_{out} \\ h & h & f_{out}^* & 0 \end{pmatrix}. \quad (7)$$

The eigen energies $E$ of the Hamiltonian (7) are found from a fourth-degree equation:

$$E^4 - A_1 E^2 - 4A_2 E - A_3 = 0, \quad (8)$$

where $A_1 = (|f_{in}|^2 + |f_{out}|^2 + h^2)$, $A_2 = h^2(|f_{in}|\cos\varphi_{in} + |f_{out}|\cos\varphi_{out})$, $A_3 = 4h^2|f_{in}||f_{out}|\cos\varphi_{in}\cos\varphi_{out} - |f_{in}|^2|f_{out}|^2$; $\varphi_{in}$ and $\varphi_{out}$ are the phase shifts between the sublattices of the inner and outer tubes, respectively (see Eq. 2).

All four solutions of the equation (8) can be obtained numerically. When the value of $h$ is small the roots of Eq. (8) do not bifurcate but only repel each other with increasing $h$. Therefore, the order of bands (from lower to higher energies) remains the same as in non-interacting nanotubes, where $h = 0$. Using this fact, let us analyze how the energies of the valence and conduction bands are shifted due to the small perturbation $h$.

First, we consider the bands of inner SWCNT.

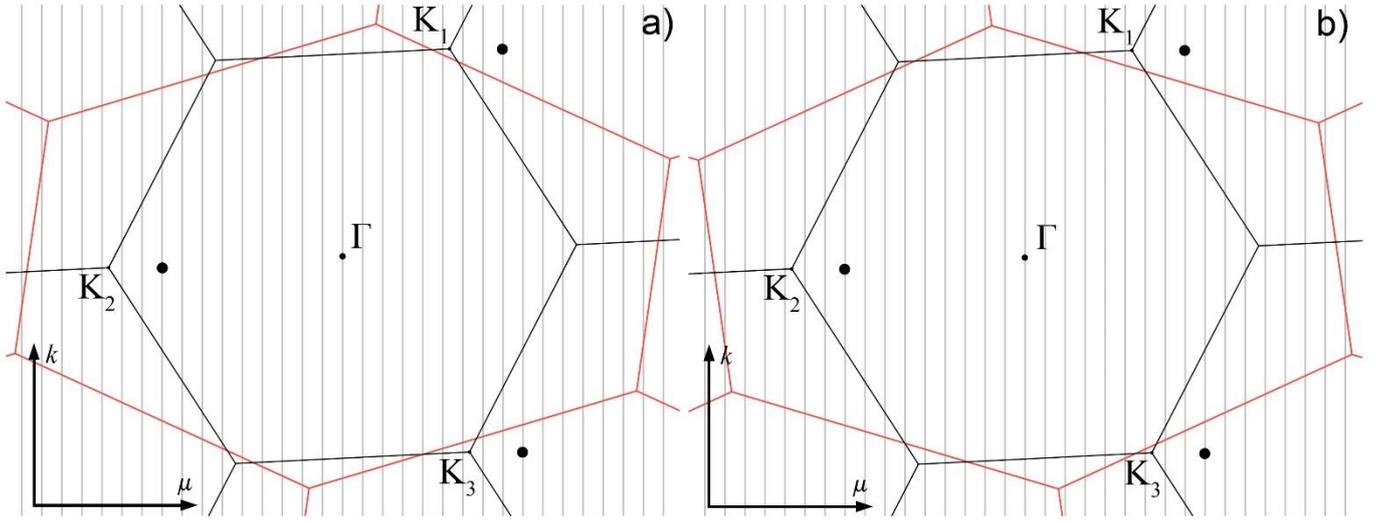

FIG. 1. The reciprocal space of the DWCNT (16,9)@(24,10). Panels (a) and (b) correspond to the cases of the same (a) and the opposite (b) handedness of the comprising nanotubes. The extended Brillouin zones of the inner and outer SWCNTs are shown in black and red, respectively. Coincident cutting lines of both tubes are denoted with gray parallel lines. With the used definition of basis vectors in direct and reciprocal spaces of SWCNT (see Appendix A), the cutting lines coincide automatically. In panel (b), the reciprocal space of the outer tube is reflected relative to the horizontal direction compared to the one shown in panel (a). Large black circles show modes with strong electronic coupling and correspond to translationally equivalent states of the inner tube. In pristine tube the energies of these states are equal, but the phases φ (see Eq. (2)) at points near $K_2$ and $K_3$ differ from the phase at the point near $K_1$ by $\pm 2\pi / 3$. For the outer tube, the same circles correspond to different states.

Suppose that the energy of the conduction band has shifted by $\Delta E_{in}^+(q)$ due to the interlayer coupling. Then, in order to find that shift, we substitute the energy $E$ as $|f_{in}| + \Delta E_{in}^+$ in Eq. (8). Further, by linearizing the equation for $\Delta E_{in}^+$ and expanding its solution in a series limited to the second order terms in $h$, we obtain

$$\Delta E_{in}^+(q) = \frac{h^2(1 + \cos \varphi_{in})(1 + \cos \varphi_{out})}{|f_{in}| - |f_{out}|} + \frac{h^2(1 + \cos \varphi_{in})(1 - \cos \varphi_{out})}{|f_{in}| + |f_{out}|}, \quad (9)$$

where all the quantities on the right-hand side also depend on the wave vector $q$. Obviously, the obtained expansion is applicable provided that $h \ll ||f_{in}| - |f_{out}||$. In addition, comparing it with the results of stationary perturbation theory,[37] one can see that the first and second terms correspond, respectively, to the intra-band and cross-band interactions between the modes.

Performing similar calculations for the VB case, one can find

$$\Delta E_{in}^-(q) = -\frac{h^2(1 - \cos \varphi_{in})(1 - \cos \varphi_{out})}{|f_{in}| - |f_{out}|} - \frac{h^2(1 - \cos \varphi_{in})(1 + \cos \varphi_{out})}{|f_{in}| + |f_{out}|}. \quad (10)$$

Accordingly, the energy shift for the direct electronic transition is obtained as

$$\Delta E_{in}(q) = \frac{2h^2(1 + \cos \varphi_{in} \cos \varphi_{out})}{|f_{in}| - |f_{out}|} + \frac{2h^2(1 - \cos \varphi_{in} \cos \varphi_{out})}{|f_{in}| + |f_{out}|}. \quad (11)$$

Let us return to a pair of strongly coupled modes with $q' = q$. As is shown in Appendix B the strong coupling takes place for all other modes equivalent to the considered ones. Accordingly, the state (or equivalent states) of one nanotube can interact with several non-equivalent ones of other nanotube[20]. As an example, Fig. 1 shows strongly coupled modes in superimposed reciprocal spaces of SWCNTs forming the DWCNT (16,9)@(24,10). In the case (a) comprising SWCNTs have the same handedness, while in the case (b) they possess the opposite one. The black circles show strongly coupled states that are chosen in the figure as follows. First, a state of the inner tube is chosen with a wave vector in the vicinity of the reciprocal space point $K_1$. After that the translationally equivalent states in vicinities of $K_2$ and $K_3$ points are also taken into account.

The three equivalent states of the inner tube (see Fig. 1) can strongly couple with three non-equivalent states of the outer tube. In turn, the latter ones can also interact with states of the inner tube which are different from the former three states. Thus, in any incommensurate DWCNT there is a branching infinite sequence of strongly coupled states. Therefore, obtaining an exact solution to the problem of DWCNT eigen energies is a very challenging task and we follow the approximation,[20] according to which the state of the inner (outer) tube is considered only at three translationally equivalent points (as is shown in Fig. 1) and its energy shift is mainly due to the interaction with three non-equivalent states of the outer (inner) tube *at the same points*. Obviously, within this approach, the energy dispersions of the outer and inner tubes have to be found one by one.

Thus, for the inner tube the shifts of CB and VB energies can be obtained as the sum over three wave vectors and the total energy shift $\Delta E_{tot}$ for the in-tube electronic transition at the point $q$ is found as

$$\Delta E_{tot}(q) = \sum_{j=1}^{3} \Delta E_{in}(q_j), \quad (12)$$

where the wave vectors are expressed as $q_1 = q$, $q_2 = q - b_1$, $q_3 = q + b_2$; $b_1$ and $b_2$ are the basis vectors of the inner tube reciprocal spaceю To simplify the calculation of the above sum, one can take into account that $|f_{in}(q_j)|$ is invariant under the translations in the reciprocal space. In addition, for all in-tube electronic transitions considered below in our paper, one can use approximate expressions (9-11). The energy shifts for the outer SWCNT are found in a similar way. To do so one only needs to swap the «in» and «out» indices in Eqs. (9-12) and use the vectors of the outer tube reciprocal space.

Note also that, as can be seen from expressions (9-11), the energy shift $\Delta E_{tot}$ in the inner tube is positive provided that the denominators of the first terms in Eqs. (9-10) satisfy the relation $|f_{in}| > |f_{out}|$. This is possible only if the strong coupling point is accidentally located in the vicinity of the $K$ point belonging to the outer tube reciprocal space. Similarly, for a transition energy in the outer tube to be positively shifted, it is necessary to satisfy the condition $|f_{out}| > |f_{in}|$, which requires proximity of the strong coupling point and the $K$ point of the inner tube reciprocal space. Analyzing the experimental data from Ref. 20, one can conclude that this situation is quite often, however, due to the presence of an additional red shift,[14,38,39] in-tube optical transitions are extremely rarely shifted toward the blue end of the spectrum. A further extensive analysis of available experimental data is mainly aimed to search for such optical transitions in DWCNTs that cannot be interpreted as in-tube one. Theoretical approach to these transitions is also developed in the next section.

## III. EXPERIMENTAL DATA ANALYSIS AND THE CONCEPT OF INTER-TUBE ELECTRONIC TRANSITIONS

Let us recall that the electronic transitions displaying themselves in the optical spectra of SWCNTs originate from extremal points of the function $f(\mu, k)$ at fixed $\mu$ values. These extrema are also called van Hove singularities.[30] Near the considered $K$ points [$\boldsymbol{K}_1 = (\boldsymbol{b}_1 - \boldsymbol{b}_2)/3$, $\boldsymbol{K}_2 = -(2\boldsymbol{b}_1 + \boldsymbol{b}_2)/3$, $\boldsymbol{K}_3 = (\boldsymbol{b}_1 + 2\boldsymbol{b}_2)/3$] the VHS coordinates can be found as $\boldsymbol{K}_j + \boldsymbol{P}$, where $\boldsymbol{P} = \left(\frac{p}{3}, \delta k\right)$, $p$ is an integer[34] and $\delta k$ is a small VHS shift along the cutting line $\mu(p)$. The value of $\delta k$ increases with the distance between the cutting line and the $K$ point. Positive and negative numbers $p$ which are multiples of three ($|p| = 3,6,9$) correspond to transitions in metallic SWCNTs ($M_{11}$, $M_{22}$, $M_{33}$). These transitions are split (except for the case of armchair tubes),[30,34] positive numbers correspond to slightly higher energies than negative ones. This splitting can also be obtained within the framework of the Hamiltonian (1).[30] Other positive and negative numbers ($|p| = 1,2,4,5,7,8$) index electronic transitions in semiconducting tubes ($S_{11}$, $S_{22}$, $S_{33}$, $S_{44}$, $S_{55}$, $S_{66}$). For the SWCNT ($n,m$) the sign of $p$ numbers is unequivocally determined by the integer constraint for the "angular" component $\mu(p) = (n - m + p)/3$ of the $\boldsymbol{q}$ vector.

In the framework of conventional NN TBA, the error in calculations of SWCNT transition energies amounts to 100-150 meV or 10-15 % which by modern standards is far from satisfactory. However, such a relative error is acceptable when calculating only a small correction $\Delta E_{tot}$ to the unperturbed energy. Therefore, using NN TBA, we calculate only the shifts $\Delta E_{tot}$, and transition energies in SWCNTs are taken from Ref. 34. By adding these two quantities we obtain optical transition energies in DWCNTs.

As was shown in Refs. 14,20, the energies of electronic transitions in DWCNTs can be shifted not only due to the interlayer coupling, but also due to the screening effect.[38,39] To take this into account in our calculations we added the constant term $\Delta_s$ to the shift $\Delta E_{tot}(p)$, which is assumed to be different for metallic and semiconducting constituent SWCNTs.

First, we have analyzed 95 optical transitions in 27 DWCNTs from Ref. 20, in which the absorption spectra of DWCNTs were measured and the historically first theory describing DWCNT optical transitions within a weak perturbation regime was developed. Note that we have excluded the DWCNT (11,7)@(21,6) from the consideration since, in our opinion, the data[20] contain typos for this nanotube. In our calculations, the hopping coefficient $\gamma$ is assumed to be 3.0 eV and 2.9 eV for semiconducting and metallic SWCNTs, respectively. On average, with these $\gamma$ values the theoretical (obtained within the framework of the NN TBA) transition energies in SWCNTs forming DWCNTs[20] turn out to be the closest to the experimental values. The remaining material coefficients $\gamma_c$ and $\Delta_s$ (the interlayer interaction strength and the screening constant) were obtained from the experimental data[20] by using the method of least squares. The resulting values are $\gamma_c \approx 933$ eV, $\Delta_s(M) = -50$ meV and $\Delta_s(S) = -60$ meV for metallic and semiconducting tubes, respectively. The obtained $\gamma_c$ value turns out to be a bit lower than the value from Ref. 20, which is apparently due to the fact that our model, in contrast to the approach,[20] explicitly takes into account the cylindrical geometry.

Using Eq. (12) we calculated the energy shifts for 94 of the 95 optical transitions,[20] the standard deviation of $\Delta E_{tot}$ is 18 meV, and the maximum deviation lies within the range from -28 meV to 47 meV. Approximately the same deviations were obtained in Ref. 20, although our expressions seem to us more accurate. In contrast to the model[20], we take into account cross-band interaction described by the second terms in Eqs. (9-10); moreover, we calculate the matrix elements using the explicit summation (6), rather than approximate integral expression (S5) from Ref. 20. The error resulting from utilization of the latter expression, according to our estimates, can reach 10%. However, in spite of the better accuracy, one of the experimentally observed transitions in the DWCNTs[20] cannot be explained in the proposed framework. Let us consider the problem in more detail.

The optical transition in the DWCNT (12,12)@(21,13) with the energy $E_{DW} = 1.76$ eV, according to Ref. 20, originates from the transition with an index $|p| = 5$ of the outer SWCNT (21,13). The transition energy $E_{out}$ in pristine nanotube is 1.91 eV and the calculations using Eq. (12) lead to a theoretical value of $E_{DW} = 1.92$ eV instead of 1.76 eV. The resulted error (160 meV) is more than 3 times larger than the maximum deviation for the other transitions.

Standard analysis shows that the outer tube interacts strongly with the inner one only near the $K'_1$ point, and the interaction near the points $K'_2$ and $K'_3$ can be completely neglected. At the coupling point, the quantities $|f_{out}| = 0.89$ eV and $|f_{in}| = 0.76$ eV are close to each other. Since $|f_{out}| > |f_{in}|$, the interlayer interaction (see Eq. (9-11)) must shift the optical transition toward higher energies, but in the experiment, we see the strong red shift.

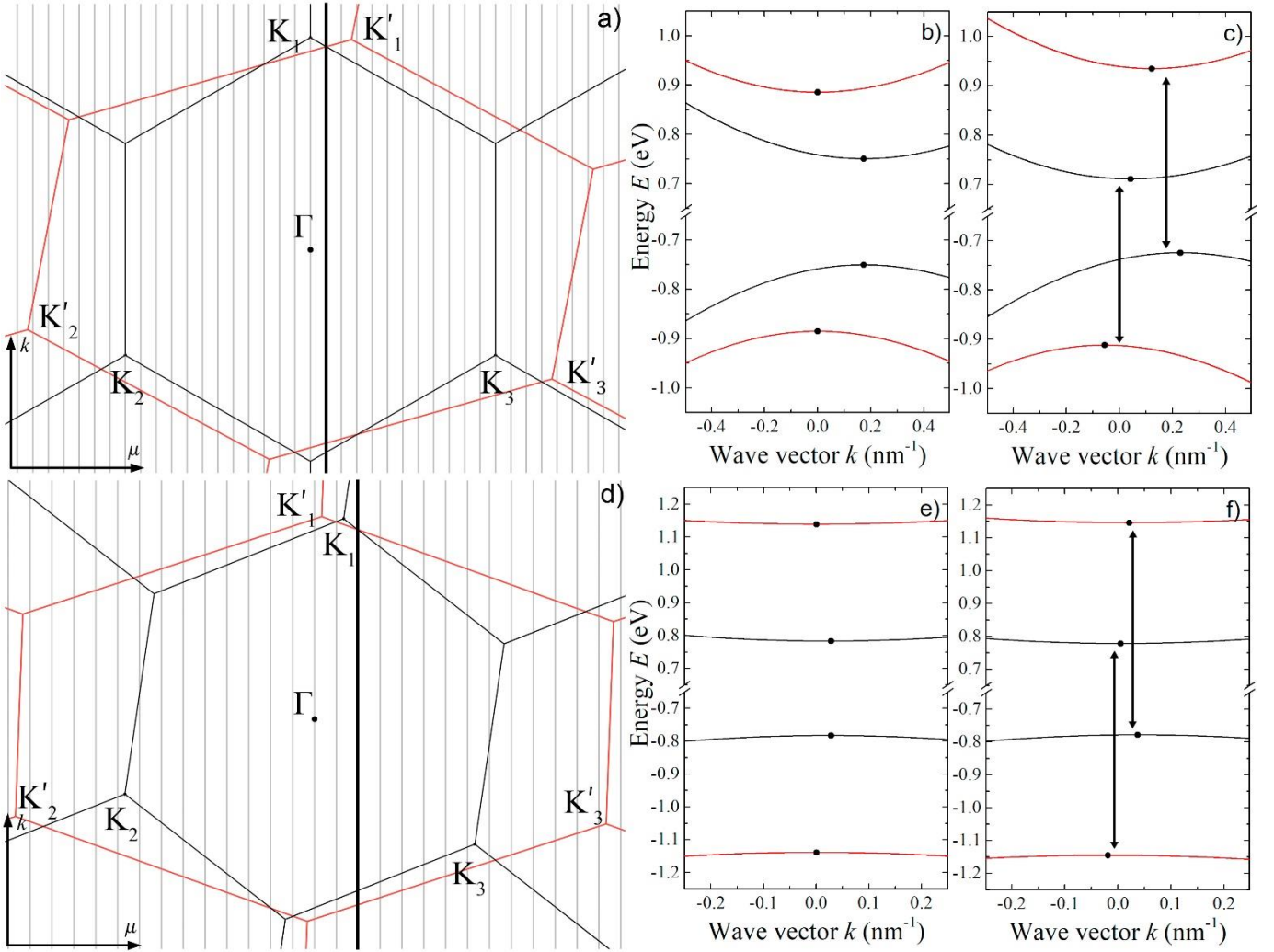

FIG. 2. Reciprocal space and dispersion relations of the DWCNT (12,12)@(21,13) (panels a, b and c) and (10,6)@(14,13) (panels d, e and f). In panels (a) and (d), the extended Brillouin zones of the inner and outer SWCNTs are shown in black and red, respectively. The cutting lines of the DWCNTs are gray; the line associated with the inter-tube transition is highlighted with black. The projections of the points $K_1'$ and $K_1$ on the highlighted lines have very close k coordinates, which corresponds to the proximity of extrema on the dispersion curves of pristine SWCNTs shown in panels (b) and (e). The interlayer coupling shifts the VHS positions and modifies the DWCNT spectra, as shown in panels (c) and (f). Dispersions for the inner and outer tubes are shown in black and red, respectively. The black circles in panels (b), (c), (e), and (f) indicate the exact VHS positions obtained within the model. Arrows denote inter-tube direct transitions in DWCNTs. In all the plots, the VHS coordinate of the outer pristine SWCNT is taken as the origin ($k=0$).

To understand the phenomenon, let us consider the scheme of superimposed extended Brillouin zones in the reciprocal space (see Fig. 2(a)). One can see that in the outer nanotube the coupling point (which is located on the cutting line with $\mu = 1$) is very close to the VHS of the inner tube. Namely, on the scale of Fig. 2(a), the projections of the points $K_1'$ and $K_1$ on this cutting line are practically indistinguishable and correspond to the observed optical transitions in the outer (with $|p| = 5$) and inner (with $p = 3$) pristine nanotubes, respectively. Thus, there is a quite rare situation when both strongly coupled modes in the DWCNT are very close to optical transitions in the spectra of the pristine SWCNTs.

Since in the inner tube the energy bands (with $\mu = 1$) are also mainly modified due to the coupling near the $K_1$ point (its contribution is about 80%), for simplicity, the further analysis of the interacting modes can be carried out within the single Hamiltonian (7). Using it, we have calculated the dispersion curves for the CBs and VBs of the DWCNT (12,12)@(21,13) near the projections of the points $K_1$ and $K_1'$ on the considered cutting line. The resulting dispersions are presented in Fig. 2(c).

The wave vector $k$ corresponding to the point $K_1'$ (see Fig. 2(a)), equals 16.86 nm$^{-1}$ and, as we earlier pointed out, on the scale of the figure the projections of the points $K_1$ and $K_1'$ on the first cutting line practically coincide. However, on the scale of Fig. 2(b) (where interlayer coupling is not taken into account), it becomes clear that the extrema of the dispersions of the inner and outer tubes do not lie exactly one above the other. The interlayer coupling leads to a rearrangement of the electronic spectrum (see Fig. 2(c)), as a result, the extrema, which previously were exactly above each other, now diverge, and the extrema of the bands that originate from different SWCNTs converge. Consequently, the wave vector with the coordinate of the CB bottom (the VB top) of the outer SWCNT approximately coincides with the one corresponding to the VB top (the CB bottom) of the inner SWCNT. In fact, calculations within our model show that the VHSs in the VB and CB of the outer tube diverge up to $\Delta k \approx 0.2$ nm$^{-1}$. The distance between the VHS in bands originating from different pristine SWCNTs becomes equal to $\approx 0.1$ nm$^{-1}$. Thus, we conclude that this location of VHSs makes possible the inter-tube optical transition and its intensity should be significantly greater than the one of the in-tube transition.

In order to compare qualitatively our theory and experimental data, let us define the effective shift in the experimental transition energy as $E_{DW} - (E_{in} + E_{out})/2$, which approximately equals –0.015 eV for the considered case. The definition implies that the CB and VB in SWCNTs are symmetrical. Such assumption is fairly reasonable considering the results of Ref. 34. Using the Hamiltonian (7) we obtain the following energies $E_{in}^+ = 0.71$ eV, $E_{in}^- = -0.725$ eV, $E_{out}^+ = 0.935$ eV, $E_{out}^- = -0.91$ eV, which values are calculated at the extrema of the dispersion curves. As can be seen in figure 2, there are two inter-tube transitions in the DWCNT spectrum with corresponding energies $\approx (E_{out}^+ - E_{in}^-)$ and $\approx (E_{in}^+ - E_{out}^-)$. A small displacement $\Delta k$ of the band extrema relative to each other (see Fig. 2c) leads to a broadening of the peaks and a slight blue shift of the transition energies[40] compared to the above approximate values. As our estimates show (within the approximation of the dipole matrix element not depending on the wave vector $\boldsymbol{q}$) for all inter-tube transitions considered in this work, the blue shift has a value less than 0.01 eV and, therefore, is not taken into account below.

Then, for one of the spectral peaks, we obtain that
$$\Delta E = (E_{out}^+ - E_{in}^-) - \frac{(E_{in}^{SW} + E_{out}^{SW})}{2} + \Delta_s \approx \Delta_s + 0.01 \text{ eV},$$
where the values $E_{in}^{SW} = 1.50$ eV and $E_{out}^{SW} = 1.77$ eV are the optical transition energies in non-interacting SWCNTs, calculated using NN TB theory. Assuming that the screening shift $\Delta_s$ is approximately the same (from -50 to -60 meV) as for in-tube transitions in metallic and semiconducting tubes, we get that the error of the proposed theory is about 35 meV. A more detailed analysis shows that the found error decreases by 15–20 meV provided that the electronic coupling near the points $K_2'$ and $K_3'$ is taken into account as well. This can be done within the above-developed perturbation theory. The second spectral peak (shifted toward the red end of the spectrum by only 40 meV) apparently was not resolved during the spectrum fitting due to its lower intensity. At the end of the next section we will demonstrate it by directly calculating optical matrix elements.

We have attempted to find in those few papers,[15,16] devoted to the experimental study of the DWCNT optical spectra, other cases that can be interpreted as inter-tube transitions. We have analyzed 23 optical transitions in 5 DWCNTs (for more details see Supplemental Material, S1) and found a particularly interesting one in the DWCNT (10,6)@(14,13) with an energy $E_{DW} = 1.985$ eV.[15] According to the authors this transition originates from the $S_{33}$ one in the outer pristine SWCNT, where its energy is 1.93 eV.[34]

Calculation using Eq. (13) gives the following energy $E_{DW} = 1.87$ eV of the DWCNT optical transition, which differs from the experimental value by more than 100 meV. Further analysis of the dispersion relations shows that the spectral line with the energy of 1.985 eV corresponds with much better accuracy to the inter-tube transition.

Fig. 2(d) shows superimposed reciprocal spaces of the tubes (10,6) and (14,13). The projections of the points $K_1$ and $K_1'$ on the cutting line with the index $\mu = 2$ practically coincide. The first and second projections correspond to the optical transitions in the inner ($|p| = 2$, $E_{in} = 1.66$ eV)[34] and outer ($|p| = 5$, $E_{out} = 2.34$ eV)[34] pristine nanotubes. The calculation of dispersion relations (see Fig. 2(f)) using the Hamiltonian (7) shows a similar spectrum rearrangement as in the case of DWCNTs (12,12)@(21,13). The distance between the extrema that originate from the bands of different pristine SWCNTs turns out to be two times smaller ($\Delta k \approx 0.02$ nm$^{-1}$) than the distance between the VHSs of CB and VB of the outer SWCNT ($\Delta k \approx 0.04$ nm$^{-1}$). The theoretical energies, calculated as $(E_{in} + E_{out})/2 + \Delta E$, are both approximately equal to 1.95 eV, which is very close to the experimental value.

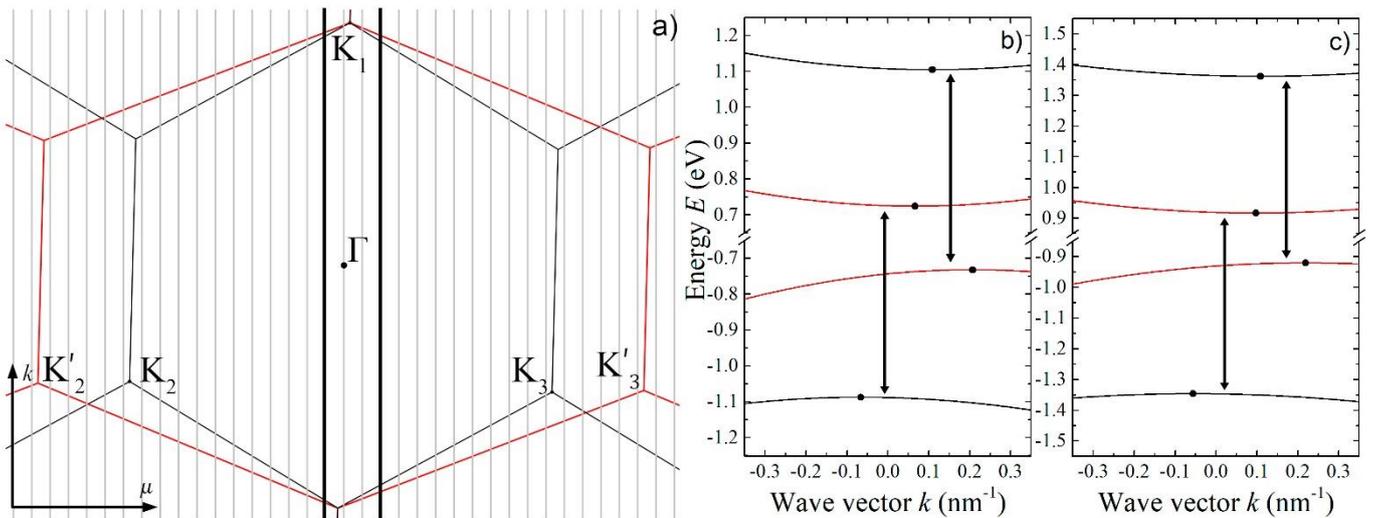

FIG. 3. Inter-tube transitions in DWCNT (12,11)@(17,16). The panel (a) demonstrates the superimposed extended Brillouin zones of the inner and outer SWCNTs which are shown in black and red, respectively. The cutting lines of the DWCNT are gray; lines $\mu = -1$ and $\mu = 2$ highlighted with black correspond to the DWCNT bands which dispersions are presented in panels (b) and (c), respectively. Dispersions for the inner and outer tubes are shown in black and red, respectively. The origin in the plots is chosen as the VHS coordinate of the inner pristine SWCNT. The VHS positions are denoted by black circles, while inter-tube transitions are denoted by arrows.

Note that, in contrast to the case of DWCNTs (12,12)@(21,13), this spectrum rearrangement does not affect the $S_{44}^{out}$ transition observed experimentally,[15] but makes possible the appearance of additional peaks in optical spectra of DWCNTs, which are not associated with transitions in pristine SWCNTs. As can be seen from the spectrum in Fig. 2(f), all four VHSs have very close coordinates, thus, we argue that in this case both the in-tube transition ($S_{44}^{out}$) and the inter-tube one (identified by the authors[15] as $S_{33}^{out}$) originate from the electronic bands corresponding to the same cutting line $\mu = 2$.

Let us also note that the authors of Ref. 15, according to their experimental data, could not unambiguously identify the indices of the outer SWCNT [(14,13) or (16,11)]. Our analysis shows that the outer tube is undoubtedly (14,13).

The first DWCNT Rayleigh spectrum, containing additional spectral peaks, which, according to Ref. 29, cannot be associated with the optical transitions in the inner and outer SWCNTs, was published only a few months ago. In the measured range from 1.5 to 2.75 eV the spectrum of DWCNT (12,11)@(17,16) has two more spectral lines with energies about 1.85 eV and 2.35 eV in addition to 5 of those originating from the pristine SWCNTs. Our theory unambiguously interprets these peaks as inter-tube transitions, which correspond to the cutting lines with the indices $\mu = -1$ and $\mu = 2$. Both values of $\Delta k$ for these cases are equal to ≈0.1 nm$^{-1}$ and are very close to $\Delta k$ values of the other inter-tube transitions considered in this paper.

According to our analysis, the first measured peak with energy 1.85 eV corresponds to transitions with theoretical energies 1.89 eV and 1.92 eV (see Fig. 3b), and the second one corresponds to transitions with energies 2.35 eV and 2.37 eV (see Fig. 3c). Unfortunately, in the work[29] neither values of transition energies (only the initial Rayleigh spectrum without its decomposition), nor specific values of energies calculated within the framework of their theory are given, so it is not possible to compare the approaches. Note that our theory also predicts the inter-tube transitions in the bands with $\mu = 0$ and $\mu = 1$ (see Fig. 3a). However, their energies are beyond the measured range and are close to 1 eV.

After analyzing the four inter-tube transitions we came to the conclusion that they are not as rare as stated in Ref. 29 and returned to the analysis of the experimental data[20] in order to find other possible transitions of this type. We checked whether the observed energies[20] and the energies of possible inter-tube transitions can match or be very close to each other. As a result, we have found out that inter-tube optical transitions can be masked in the spectra of 4 DWCNTs (see Table 1). Calculated DWCNT electronic dispersions explaining the origin of the transitions in Table I are shown in Supplemental Material, S2. It is interesting to note that all the DWCNTs, in which inter-tube transitions are possible, are composed of the tubes with the same handedness.

In the following section we propose an approach to calculate dipole matrix elements. Within its framework, we demonstrate that in considered cases, the intensity of the inter-tube transition can be of the same order as the intensity of conventional in-tube one.

## IV. OPTICAL MATRIX ELEMENTS FOR INTER-TUBE TRANSITIONS

Previously in the Article, we justified the existence of inter-tube transitions in DWCNTs by the fact that without introducing this concept, it turned out to be impossible to explain the measured energies of some optical transitions. We have also found the specific band rearrangement corresponding to the case. However, to unequivocally prove the existence of inter-tube optical transitions we need to consider the Hamiltonian $H_{opt}$ describing the coupling with the electromagnetic field. Within the dipole approximation this Hamiltonian can be written as[41]

$$H_{opt} = \zeta \eta (\boldsymbol{P} \cdot \nabla), \quad (13)$$

where $\zeta = -i \frac{e\hbar}{m_0} \sqrt{\frac{\hbar}{2V_0 \varepsilon_0 \omega}}$, $e$ and $m_o$ are charge and mass of the electron, $\varepsilon_0$ is dielectric constant, $V_0$ is the DWCNT volume, quantity $\eta = \sqrt{N_p}$ corresponds to the case of photon absorption, while $\eta = \sqrt{N_p + 1}$ corresponds to its emission, $N_p$, $\omega$, and $\boldsymbol{P}$ are the photon number, frequency and polarization, respectively.

An electron transition from the state $|i\rangle$ to the state $|f\rangle$ is allowed if the matrix element $\langle f|H_{opt}|i\rangle$ does not vanish. However, instead of calculating matrix elements directly let us approach this problem by, first, rewriting the operator (13) in the matrix representation. Since for all considered inter-tube transitions the interaction between tubes near the $K_1$ point is significantly higher, we neglect the interaction near $K_2$ and $K_3$ points. This allows us to choose the same four Bloch WFs we previously used to construct the effective Hamiltonian (7) as a new representation basis. We also consider the light polarization $\boldsymbol{P}$ along DWCNT axis ($\boldsymbol{P} = \langle 0,0,1 \rangle$). This choice corresponds to the discussed optical transitions which preserve the value of the wave number $\mu$. Then, within the framework of the nearest neighbor tight-binding approximation, we rewrite the Hamiltonian (13) as

TABLE I. Possible inter-tube optical transitions in the DWCNTs[20].

| DWCNT | Cutting line, $\mu$ | Energy, eV | Distance between VHSs for inter-tube transitions, nm$^{-1}$ |
|---|---|---|---|
| (13, 8)@(16, 15) | 2 | 1.39, 1.39 | 0.05, 0.13 |
| (14, 9)@(17, 16) | 3 | 2.55, 2.57 | 0.12, 0.11 |
| (18, 5)@(27, 5) | 6 | 2.10, 2.10 | 0.05, 0.01 |
| (13, 2)@(21, 3) | 4 | 1.74, 1.77 | 0.17, 0.12 |

$$H_{opt} = i\zeta\eta \begin{pmatrix} 0 & g_{in} & v & v \\ g_{in}^* & 0 & v & v \\ v & v & 0 & g_{out} \\ v & v & g_{out}^* & 0 \end{pmatrix}, \quad (14)$$

where, following Refs. 42 and 43, $g_{in}$ and $g_{out}$ can be compactly written as

$$g_{in} = \frac{M(\boldsymbol{a}_0)}{a_0 \gamma} \frac{\partial f_{in}}{\partial k}, g_{out} = \frac{M(\boldsymbol{a}_0)}{a_0 \gamma} \frac{\partial f_{out}}{\partial k},$$

and $M(\boldsymbol{a}_0) = \int_{S_0} \phi^*(\boldsymbol{r} - \boldsymbol{a}_0)\phi_1(\boldsymbol{r}) \, dS$; $\phi(\boldsymbol{r} - \boldsymbol{R})$ is atomic $p$-orbital, localized near the site $\boldsymbol{R}$; $\boldsymbol{a}_0$ is the vector connecting the nearest carbon atoms. Since the function $\phi(\boldsymbol{r})$ is a $p$-orbital, oriented perpendicularly to the tube axis, it can be written as $\phi(\boldsymbol{r}) = r_\perp \xi(r)$, where $r_\perp$ is the perpendicular to the z-axis projection of the radius vector, and $\xi(r)$ is some spherically symmetric real function of $r = |\boldsymbol{r}|$, localized near the origin. Accordingly, $\phi_1(\boldsymbol{r})$ denotes the function $r_\perp \xi'(r)$, where the prime means differentiation with respect to $r$. Therefore, in $g_{in}$ and $g_{out}$ the quantity $M$ only depends on the length of the vector $\boldsymbol{a}_0$.

Inter-tube matrix elements $v$ in Eq. (15) are obtained as

$$v = \frac{-1}{\sqrt{N_1 N_2}} \sum_{j,l} \frac{(z_j - z_l')}{|\boldsymbol{R}_j - \boldsymbol{R}_l'|} \sin[k \cdot (z_j - z_l')] \times \\ \times \cos[\mu \cdot (\theta_j - \theta_l')] M(\boldsymbol{R}_j - \boldsymbol{R}_l'),$$

where, as in Eq. (6), the primed and unprimed coordinates specify the positions of atoms in one of the sublattices of the inner and outer tubes. The equality of the elements $v$ with different indices is prompted by approximately random relative arrangement of the atoms in the inner and outer layers. A more rigorous justification can be carried out in the same way we used to derive Eq. (6) and which is discussed in Appendix B. For the same reasons, in the expression for $v$ the only remaining part of the factor $(z_j - z_l') \exp[-i\boldsymbol{q} \cdot (\boldsymbol{R}_j - \boldsymbol{R}_l')]$ is proportional to the term $(z_j - z_l')\sin[k \cdot (z_j - z_l')]\cos[\mu \cdot (\theta_j - \theta_l')]$, which is even with respect to independent permutations $z_j \leftrightarrow z_l'$ and $\theta_j \leftrightarrow \theta_l'$. Deriving the Hamiltonian (14), we have also utilized the approximation,[44] according to which

$$\left\langle \phi(\boldsymbol{r} - \boldsymbol{R}_j) \left| \frac{\partial}{\partial z} \right| \phi(\boldsymbol{r} - \boldsymbol{R}_l) \right\rangle \approx \\ \frac{z_j - z_l}{|\boldsymbol{R}_j - \boldsymbol{R}_l|} \int_{V_0} \phi^*(\boldsymbol{r} - \boldsymbol{R}_j + \boldsymbol{R}_l)\phi_1(\boldsymbol{r}) \, dV.$$

The uncertainty in the approximation when $\boldsymbol{R}_j - \boldsymbol{R}_l = 0$ requires a separate analysis, which shows that in this case the element $\left\langle \phi(\boldsymbol{r} - \boldsymbol{R}_j) \left| \frac{\partial}{\partial z} \right| \phi(\boldsymbol{r} - \boldsymbol{R}_l) \right\rangle$ vanishes.

The Hamiltonian (7) simultaneously determines the energies $E_i$ and the corresponding normalized eigenvectors $\boldsymbol{C}_i$ of four electron states with the same wavevector $\boldsymbol{q}$. Within the framework of the developed approach the dipole matrix element of the electron transition between $i$-th and $j$-th states can be calculated as:

$$d_{i,j} = \boldsymbol{C}_j^* \boldsymbol{H}_{opt} \boldsymbol{C}_i. \quad (15)$$

By substituting the eigenvectors of the matrix (7) into Eq. (15) when the interaction between tubes is neglected ($h = 0$), one can see that the dipole matrix elements for individual SWCNTs coincide with the ones obtained in the previous works,[41,43] and when $h = 0$ the inter-tube transitions are not allowed.

As our numerical estimates show, for all inter-tube transitions considered in the previous section the electron states are strongly mixed due to Van-der-Waals interaction; namely, the lengths of the vectors $\boldsymbol{C}_i$ and $\boldsymbol{C}_j$ projected onto the subspaces corresponding to the inner and outer tubes turn out to be close to each other. Thus, even when $v \ll g_\alpha$ ($\alpha = in, out$), the optical matrix elements for inter-tube transitions do not vanish. However, accurate calculations of the quantity $d_{i,j}$ may prove difficult, since there are no estimates of the function $M(\boldsymbol{R})$ in the literature.

Let us note, that in case of $p$-orbitals function $M(\boldsymbol{R})$ as well as the overlap integral $\int_{V_0} \phi^*(\boldsymbol{r} - \boldsymbol{R})\phi(\boldsymbol{r}) \, dV$ depends on the value of the parallel and perpendicular projection of $\boldsymbol{R}$ to the DWNCT radius. However, since the interlayer coupling is well described by the function $u = \gamma_c \exp(-r/\lambda)$, where $r = |\boldsymbol{R}|$, we can roughly estimate the integral $\int_{V_0} \phi^*(\boldsymbol{r} - \boldsymbol{R})\phi'(\boldsymbol{r}) \, dV$ as $\frac{\partial u}{\partial r}$ (such estimate is valid for spherically symmetric localized states). Then, as our numerical estimates show, when the intra-layer overlap integral $M(\boldsymbol{a}_0)$ has an order of 0.5 - 1.0 eV nm$^{-1}$, the dipole matrix elements of the inter-tube and in-tube transitions turn out to be very close.

Using this estimate, it is interesting to return to the discussion of the inter-tube transitions in DWCNT (12,12)@(21,13) considered in the previous section. For both transitions with close energies $\approx (E_{out}^+ - E_{in}^-)$ and $\approx (E_{in}^+ - E_{out}^-)$ the matrix elements and eigenvectors of the states involved have relatively weak dependence on the wave vector. However, since these transitions involve different pairs of eigenvectors of the Hamiltonian (7), the dipole matrix elements can be very different.

The intensity of a peak in absorption spectrum is proportional to $|d_{i,j}|^2/(E_i - E_j)^2$.[41] Then using the estimates for $M(\boldsymbol{a}_0)$, one can obtain that the intensity of the transition with energy $\approx (E_{out}^+ - E_{in}^-)$ is higher by 30%, when $M(\boldsymbol{a}_0) = 1$ eV nm$^{-1}$, and only by 10%, when $M(\boldsymbol{a}_0) = 0.5$ eV nm$^{-1}$, which is consistent with the experimental data[20]. Similar calculations for the rest of the inter-tube transitions considered in this paper are not presented here, since for these transitions it is difficult to compare theoretical results with the available experimental data.

## V. DISCUSSION

In this work, we developed the theory of inter-tube optical transitions in DWCNTs. As the first step we have revised the NN TBA which is proven to be a very powerful tool in calculating the band structure of low-dimensional systems.[45,46] Unlike Ref. 20, we have not applied the perturbation theory from the very beginning and first considered the effective 4x4 Hamiltonian for strongly coupled states of the outer and inner tubes. Constructing this Hamiltonian, we explicitly have taken into account the incommensurability of SWCNTs in a DWCNT and obtained the more accurate expression (8) to calculate the energies of the coupled states. Its linearization leads to the partially known results. If we exclude the second (less significant) terms corresponding to cross-band interaction in the approximate expressions (9-10) describing the shifts in band dispersions, the formulas become equivalent (as we have shown) to the implicit ones obtained in Ref. 20. Our approach is more accurate for one more reason. We have refused to use the approximate expression[20] for calculating the matrix elements. Replacing the sum (6), which can be calculated on any personal computer in a few seconds, with an approximate integral expression for a slight increase in the efficiency of calculations, it seems unimportant.

As in our work, the approach[28] considers the effective 4x4 Hamiltonian. This Hamiltonian is based on the linear expansion of the function $f(\boldsymbol{q})$ (see Eq. (1)) in the vicinity of the points $K$ and $K'$.[30,32] The matrix elements describing the interaction between the sublattices of different tubes are also expanded into series near these points. In our opinion both expansions are unnecessary; they decrease the accuracy of the theory.[28] We have decided that it is better to consider only one small parameter - the magnitude of the interaction between the tubes. Indirect evidence that the theory proposed is more accurate is that we were able to detect several inter-tube transitions in the previously published data[15,20] on the optical spectra of DWCNTs. In addition, we fully agree with the authors of Ref. 20 that some state of one tube can be strongly coupled with several nonequivalent states of the other tube. Accordingly, it is necessary to take into account all the interacting states, and in general case this is impossible within the framework of the single 4x4 Hamiltonian. In general case, the 8x8 Hamiltonian may be useful, and we will deduce it elsewhere. Here, for simplicity, we successively apply three copies of equations based on the rather simple Hamiltonian (7).

The phenomenon we call the inter-tube transition was discovered a few months ago in the DWCNT (12,11)@(17,16)[29] and the authors considered it as absolutely unique, requiring complete altering of the band structure. According to Ref. 29 such transitions take place only in those DWCNTs, both layers of which have close chirality angles. In our opinion, the geometric selection rules permitting the inter-tube transitions are not so rigorous. Both inter-tube and in-tube optical transitions originate from those in SWCNTs; however, the inter-tube transition is genetically related with two different transitions occurring in the inner and outer pristine nanotubes. In the scheme of superimposed extended Brillouin zones both transitions in the nanotubes should be characterized by the same cutting line $\mu$ and have very close values of the one-dimensional wave vector $k$ (very close VHS positions on this cutting line). Then, due to the inter-tube coupling, these positions can practically coincide, which, in our opinion, makes the transition between the bands of different tubes possible.

As we mentioned the greater the distance between the considered cutting line and the $K$ point is, the more the VHS shifts along the cutting line. Therefore, the inter-tube transition is more probable to occur when the strong coupling point is near both $K$ points of the outer and inner SWCNTs. We also note that electronic bands are periodic in reciprocal space of pristine SWCNTs. Thus, the difference $\Delta k$ between the projections of $K$ points on the considered cutting line should be significantly smaller than the reciprocal space periods of the both tubes.

The above selection rules are purely geometric and can be easily applied separately without energy calculations. Even though within the proposed theory we cannot unequivocally establish the threshold value of $\Delta k$ permitting an inter-tube transition, the developed approach easily allows one to determine which transition (inter-tube or in-tube one) is more likely in a particular case. Let us also emphasize that the inter-tube transitions in the framework of our theory are described in almost the same way as the in-tube ones. If the geometrical analysis shows the possibility of inter-tube transition, in addition to the shifts of the band energies, it is necessary to determine new VHS positions and analyze their relative location.

Within the proposed theory, in addition to the case of DWCNT (12,11)@(17,16),[29] we have found two more examples where the experimental data cannot be explained by only in-tube transitions. The examples violate the selection rules,[28,29] since in both cases the difference of the chirality angles is quite significant. However, the found examples satisfy the selection rules presented in this work. In addition, analyzing the data,[20] we have revealed four probable inter-tube transitions that are close in energies with the in-tube transitions in the same DWCNTs.

In this regard, it is of great interest to re-analyze the experimental data[15,20] since for DWCNTs studied there the theoretical energies of in-tube and inter-tube transitions are well distinguishable. Unfortunately, the optical spectra of DWCNTs considered in these works are insufficiently represented. The only published Rayleigh spectrum (which according to our theory contains an inter-tube transition) is given in low resolution for the DWCNT (10,6)@(14,13).[15] In this spectrum, the maximum with an energy about 1.9–2.0 eV corresponds exactly to the wide region where, along with the in-tube transition ($S_{33}^{out}$), the inter-tube transition is located. But because of low resolution it seems almost impossible to fit the spectrum[15] and conduct a more thorough analysis.

All existing methods for fitting optical spectra of DWCNTs use the number of transitions as an input parameter which is equal to the number of transitions in pristine SWCNTs in the measured spectral range. According to Ref. 20, when fitting experimental data, a spectral line can be shifted from -200 to +50 meV with respect to its initial position and it is assumed there are no new transitions due to the interlayer coupling. Our results show that in cases satisfying the obtained geometric selection rules, additional inter-tube transitions must be taken into account. As the initial energy of such a transition, one should take the half-sum of the energies of the corresponding transitions in SWCNTs. In the most general case, the inter-tube transition is a doublet due to the symmetry breaking with respect to the electron-hole permutation. Analysis within the framework of our theory can predict the splitting and values of dipole matrix elements for possible inter-tube transitions. In subsequent works, testing the proposed theory, it would be interesting to study how the fit quality of experimental optical spectra changes when the theoretical predictions regarding the number of allowed transitions are taken into account.

## VI. CONCLUSION

In conclusion, we have developed the theory describing the band structure of incommensurate DWCNTs and introduced the concept of inter-tube electronic transitions.

They are possible in some DWCNTs with the specific geometry of superimposed extended Brillouin zones corresponding to constituent SWCNTs. The obtained selection rules permitting inter-tube transitions are purely geometric and represent a simple guideline allowing one to determine which transitions in which DWCNTs are possible even without direct calculations of the band structure. The proposed theory together with Rayleigh spectroscopy can be used as a powerful tool in the structural identification of DWCNTs including such a parameter as relative handedness of the layers. Our approach can be generalized for the case of double-walled boron nitride nanotubes[47,48] and we suppose it could also provide insight into the energy shifts of photoluminescence transitions observed in experiments on single-walled carbon nanotubes with organic molecules wrapped around them.[49]

**Acknowledgements**

D. Ch., and S. R. acknowledge financial support from the Russian Foundation for Basic Research (grant № 18-29-19043 mk). We thank C. Jin for helpful comments on Ref. 20 as well as D. Levshov and A. Myasnikova for useful discussions.

## APPENDIX A. DESCRIPTION OF THE DIRECT AND RECIPROCAL SPACES IN SWCNTS

Considering SWCNT translational and rotational symmetry, it is convenient to project graphene basis translations $\boldsymbol{a}_1$ and $\boldsymbol{a}_2$ on the surface of a nanotube[35,36] and express the components of these translations along its perimeter and axis. In contrast to the conventional approach, here, the first component of the translations is dimensionless and found as a fraction of the SWCNT perimeter. For a SWCNT with the chiral indices ($n$, $m$) one can obtain:

$$\boldsymbol{a}_1 = \left( \frac{(2n+m)\pi}{(n^2+m^2+nm)}, \frac{3m \cdot a_0}{2\sqrt{n^2+m^2+nm}} \right), \quad (A1)$$

$$\boldsymbol{a}_2 = \left( \frac{(2m+n)\pi}{(n^2+m^2+nm)}, \frac{-3n \cdot a_0}{2\sqrt{n^2+m^2+nm}} \right), \quad (A2)$$

where $a_0 = 0.142$ nm is the distance between the nearest sites of carbon atoms. Using the condition $a_i b_j = 2\pi \delta_{ij}$, where $\delta_{ij}$ is the Kronecker delta, one can easily obtain the components of the reciprocal lattice vectors $\boldsymbol{b}_1$ and $\boldsymbol{b}_2$:

$$\boldsymbol{b}_1 = \left( n, \frac{2\pi}{3a_0} \frac{2m+n}{\sqrt{n^2+m^2+nm}} \right), \quad (A3)$$

$$\boldsymbol{b}_2 = \left( m, -\frac{2\pi}{3a_0} \frac{2n+m}{\sqrt{n^2+m^2+nm}} \right). \quad (A4)$$

The used definition of the SWCNT basis vectors is convenient in the way that it leads to the automatic matching of the cutting lines in the reciprocal spaces of the outer and inner tubes and the coefficients corresponding to the reciprocal space stretching/compression used in Refs. 20,28 become unnecessary. This is because the first component of the vectors (A3-A4) measures the distance from the origin in units equal to the distance between the nearest cutting lines.

## APPENDIX B. ANALYSIS OF THE INTERLAYER COUPLING MATRIX ELEMENTS

The calculation of the elements (5) can be significantly simplified for incommensurate DWCNTs, which are the vast majority of possible DWCNTs. In such nanotubes, the ratio between the periods of the inner and outer tubes is an irrational number. Therefore, any relative shift of nanotubes in an infinitely long DWCNT can be expressed with arbitrarily good accuracy as an integer linear combination of the periods of the inner and outer SWCNTs. Any translations leave the SWCNT structure invariant; therefore, the sum (5) should also remain the same.

Unlike the relative shifts, relative rotations of nanotubes in DWCNT preserve the value of the sum (5) only approximately. Indeed, as is well known,[35,36] SWCNT ($n$, $m$) has a screw $Q$-fold axis, where $Q = \frac{2(n^2+m^2+n \cdot m)}{GCD(2n+m, 2m+n)}$ and $GCD(x, y)$ denotes the greatest common divisor of the integers $x$ and $y$. Suppose that the inner and outer nanotubes have $Q_{in}$-fold and $Q_{out}$-fold axes, respectively. Then, taking into account the translational invariance of Eq. (5), it is easy to see that Eq. (5) should also be invariant with respect to the relative rotation of the SWCNTs by the angle $\delta = 2\pi\, GCD(Q_{in}, Q_{out})/(Q_{in}Q_{out})$. In real DWCNTs, the angles are very small, for example, the maximal angle $\delta$ for all DWCNTs considered in Ref. 20 is approximately $2 \cdot 10^{-3}$ rad.

The analysis performed shows that the DWCNT symmetry with respect to relative rotations of constituent nanotubes is almost continuous, and in a sufficiently long DWCNT the pair correlation function of the atomic coordinates of inner and outer layers practically does not change under their relative rotations. Therefore, the matrix elements $V_{\alpha,\beta}$ can be considered as independent on the $\alpha$ and $\beta$ indices.

Sum (5) also must be nearly invariant with respect to a specific transformation: $\boldsymbol{R}_j \to -\boldsymbol{R}_j$ and $\boldsymbol{R}'_l \to -\boldsymbol{R}'_l$, which is equivalent to some relative translation and rotation. Therefore, the matrix element $V_{\alpha,\beta}$ should be approximately real. Indeed, since incommensurate sublattices can be relatively shifted, it is always possible to match the coordinates of a pair of atoms $\boldsymbol{R}_l = \boldsymbol{R}'_k$ and to choose in this point a common origin. The latter fact makes another interesting property of the elements (5) obvious. This sum is invariant with respect to the following substitutions:

$$\boldsymbol{q} \to \boldsymbol{q} + \boldsymbol{Q}, \boldsymbol{q}' \to \boldsymbol{q}' + \boldsymbol{Q}', \quad (B1)$$

where $\boldsymbol{Q}$ and $\boldsymbol{Q}'$ are arbitrary translations in the reciprocal space of the inner and outer nanotubes, respectively. It is interesting to note that the quantity $f(\boldsymbol{q})$ has no such property, and being translated in the reciprocal space it either remains invariant or changes the phase by $\pm \frac{2\pi}{3}$; nevertheless, this fact does not violate the translation invariance of the eigenvalues of the Hamiltonian (1).

Our calculations show that the absolute values of matrix elements $V_{\alpha,\beta}$ differ from each other by no more than 2%, and the arguments of these practically real numbers vary within $\pm\, 0.01$. In particular, for all DWCNTs considered in this paper, the maximum error occurs only for the (11,11)@(22,9) nanotube. For all other cases, the error due to utilization of approximate Eq. (6) turns out to be tens or even hundreds of times smaller.

# Supplemental Material

## Band structure and inter-tube optical transitions in double-walled carbon nanotubes

**D.V. Chalin and S.B. Rochal**[*]

Faculty of Physics, Southern Federal University, 5 Zorge str., 344090, Rostov-on-Don, Russia

---

[*] Corresponding author. Tel: +7(863)2184000, ext. 11416. E-mail: rochal_s@yahoo.fr.

**S1. Analysis of experimental data [1, 2] within the framework of the proposed theory**

In the papers [1, 2] the Rayleigh spectra of 5 DWCNTs were measured. In these spectra authors identified 23 electronic transitions, among which we detected an inter-tube one in the DWCNTs (10,6)@(14,13) [1]. To calculate the energies of the remaining 22 transitions, we applied the developed approach. In these calculations we used the same material constants as for the analysis of experimental data [3]. The resulting standard deviation is 23 meV, and the maximum deviations lie within the range from – 29 meV to 41 meV. The handedness of the DWCNTs was determined by minimizing the standard deviation. As can be seen, our theory is in good agreement with the data [1, 2] but the deviations are slightly higher than in the case of data [3]. We associate the latter with the fact that the DWCNTs [1, 2] contain adsorbed water molecules [4], which can affect the electronic properties of the DWCNTs.

*Table S1. Comparison of the experimental data [1, 2] with the theoretical calculations.*

| № | DWCNT | Transition | $E_{DW}$ | $E_{SW}$[5] | $\Delta E^{exp}$ | $\Delta E^{calc}$ | $|\Delta E^{exp} - \Delta E^{calc}|$ | Handedness |
|---|---|---|---|---|---|---|---|---|
| 1 | (7,6)@(16,6) | $S_{22}^{in}$ | 1.82 | 1.93 | -0.11 | -0.132 | 0.022 | -1 |
| 2 | (7,6)@(16,6) | $S_{33}^{out}$ | 2.09 | 2.14 | -0.05 | -0.067 | 0.017 | -1 |
| 3 | (10,6)@(14,13) | $S_{22}^{in}$ | 1.60 | 1.65 | -0.05 | -0.081 | 0.031 | 1 |
| 4 | (10,6)@(14,13) | $S_{44}^{out}$ | 2.335 | 2.34 | -0.005 | -0.046 | 0.041 | 1 |
| 5 | (14,1)@(15,12) | $S_{22}^{in}$ | 1.55 | 1.67 | -0.12 | -0.100 | 0.020 | -1 |
| 6 | (14,1)@(15,12) | $S_{33}^{in}$ | 2.44 | 2.51 | -0.07 | -0.102 | 0.032 | -1 |
| 7 | (14,1)@(15,12) | $M_{22}^{out-}$ | 2.61 | 2.67 | -0.06 | -0.056 | 0.004 | -1 |
| 8 | (14,1)@(15,12) | $M_{11}^{out-}$ | 1.44 | 1.47 | -0.03 | -0.051 | 0.021 | -1 |
| 9 | (15,9)@(22,12) | $M_{11}^{in-}$ | 1.53 | 1.59 | -0.06 | -0.064 | 0.004 | -1 |
| 10 | (15,9)@(22,12) | $M_{11}^{in+}$ | 1.57 | 1.67 | -0.10 | -0.085 | 0.015 | -1 |
| 11 | (15,9)@(22,12) | $M_{22}^{in-}$ | 2.64 | 2.79 | -0.15 | -0.137 | 0.013 | -1 |
| 12 | (15,9)@(22,12) | $M_{22}^{in+}$ | 2.93 | 3.08 | -0.15 | -0.110 | 0.040 | -1 |
| 13 | (15,9)@(22,12) | $S_{33}^{out}$ | 1.53 | 1.57 | -0.04 | -0.062 | 0.022 | -1 |
| 14 | (15,9)@(22,12) | $S_{44}^{out}$ | 1.91 | 2.00 | -0.09 | -0.062 | 0.028 | -1 |
| 15 | (15,9)@(22,12) | $S_{55}^{out}$ | 2.38 | 2.48 | -0.10 | -0.073 | 0.027 | -1 |
| 16 | (15,9)@(22,12) | $S_{66}^{out}$ | 2.93 | 2.97 | -0.04 | -0.062 | 0.022 | -1 |
| 17 | (16,12)@(27,10) | $S_{33}^{in}$ | 1.81 | 1.88 | -0.07 | -0.10 | 0.030 | 1 |
| 18 | (16,12)@(27,10) | $S_{44}^{in}$ | 2.15 | 2.34 | -0.19 | -0.180 | 0.010 | 1 |
| 19 | (16,12)@(27,10) | $S_{55}^{in}$ | 2.75 | 2.90 | -0.15 | -0.133 | 0.017 | 1 |
| 20 | (16,12)@(27,10) | $S_{33}^{out}$ | 1.44 | 1.51 | -0.07 | -0.063 | 0.007 | 1 |
| 21 | (16,12)@(27,10) | $S_{44}^{out}$ | 1.67 | 1.72 | -0.05 | -0.070 | 0.020 | 1 |
| 22 | (16,12)@(27,10) | $S_{55}^{out}$ | 2.44 | 2.49 | -0.05 | -0.063 | 0.013 | 1 |

**S2. The band rearrangement in the DWCNTs [3]**

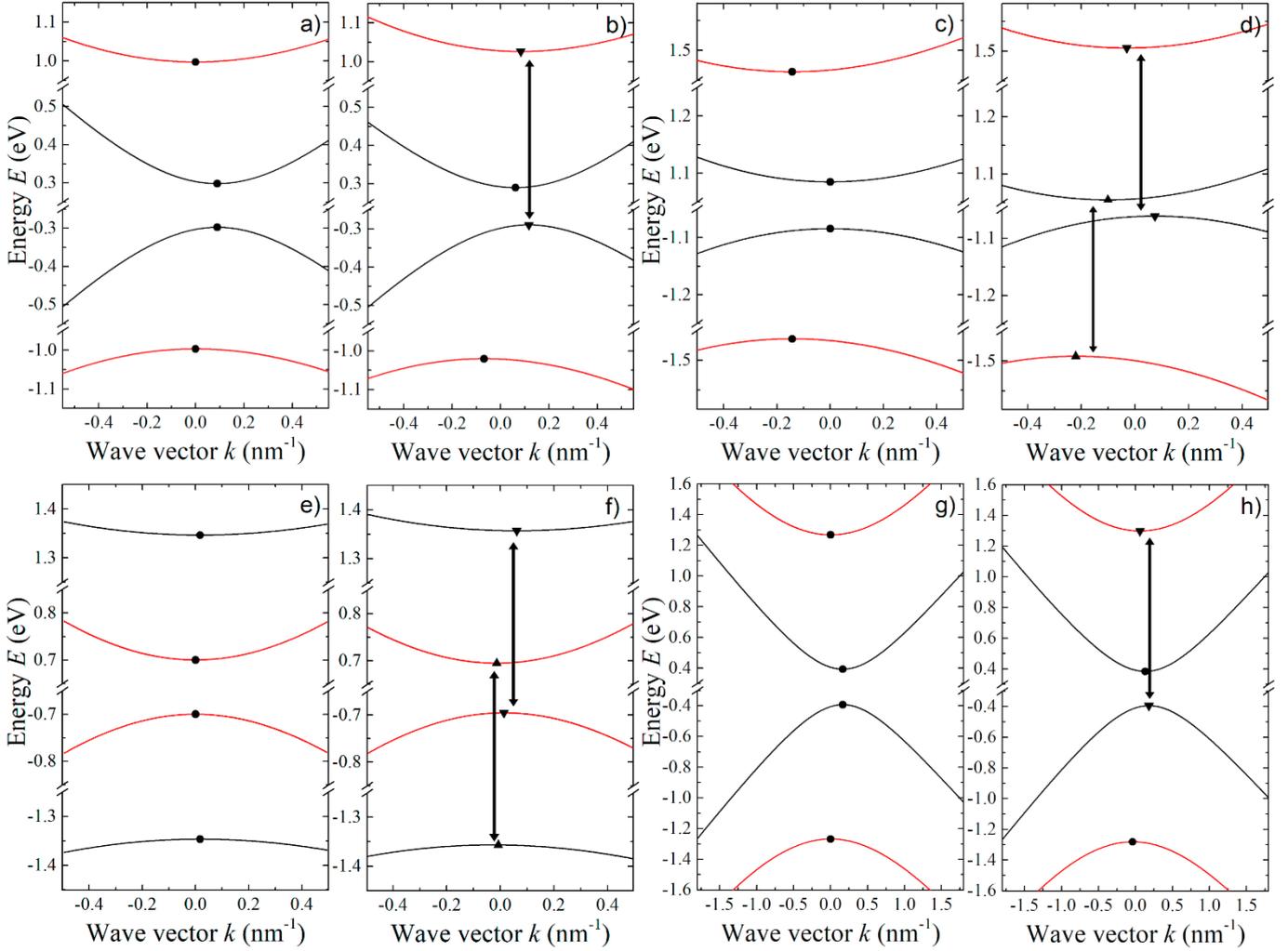

*Fig. S1. Calculated electron dispersions of DWCNTs (13, 8)@(16, 15) [panels (a),(b)], (14, 9)@(17, 16) [panels (c),(d)], (18, 5)@(27, 5) [panels (e),(f)] and (13, 2)@(21, 3) [panels (g),(h)]. For all considered DWCNTs the inter-tube transition arises due to strong coupling near the $K_1$ point. The dispersions of the inner and outer tubes are shown in black and red, respectively. In panels (a,c,e,g) the bands of non-interacting SWCNTs comprising the considered DWCNTs are shown. Panels (b,d,f,h) demonstrate the bands modification due to the coupling. The VHS positions are indicated with triangles and circles and possible inter-tube transitions are shown by arrows. In all the plots except for the (c) and (d) the VHS coordinate of the outer pristine SWCNT is taken as the origin (k=0).*